\documentstyle[preprint,aps,eqsecnum]{revtex}
\begin{document}
\draft
\baselineskip 0.6cm

\title{Effects of Diurnal Modulation in Direct \\
Cold Dark Matter Searches. \\  The Experiment in Sierra Grande}
\author{D.\ E.\  Di Gregorio, A.\ O.\ Gattone, H.\ Huck, A.\ O.\
Macchiavelli and S.\ Gil}
\address{Departamento de F\'{\i}sica, Comisi\'on Nacional de Energ\'{\i}a
At\'omica \\Av. del Libertador 8250, 1429 Buenos Aires, Argentina}
\author{J.\  Collar and F.\ T.\ Avignone III }
\address{Departament of Physics and Astronomy, University of South
Carolina,\\ Columbia, SC 29208, USA.}

\date{\today}
\maketitle
\begin{abstract}
Description and some details about the experiment to be conducted at
Sierra Grande, Argentina are presented. The potential advantages of
using the Earth as an absorber to produce diurnal modulation effects in
cold dark matter searches are given.
\end{abstract}

\section{Introduction}

By now there is compelling evidence that most of the mass in the
Universe is dark~\cite{roulet}.  The hardest evidence comes
from:\begin{enumerate}\item the flat rotation curves at galaxy
scales (measured in spiral galaxies)~\cite{faber}, \item the
anomalous line--of--sight velocities, greater than the escape velocity,
at the level of clusters of galaxies,~\cite{steigman} \item the
formation of structures in the early universe and the growth of
perturbations~\cite{olive}, and \item gravitational lensing (Einstein
rings) of very distant systems.~\cite{tyson} \end{enumerate}

{}From the measurements of mass to light ratios at different length scales a
few conclusions are drawn:  \begin{enumerate} \item The density of luminous
matter, $\rho_{lum}$, is of the order of 1\% of the critical density; {\em
i.e.} $\Omega_{lum}\equiv \rho_{lum}/\rho_c\approx 0.01$ ($\rho_c\approx
2.6h_{50}^2$ keV cm$^{-3}$), \item up to the largest scales measured,
namely that of clusters of galaxies, the dark matter density (non--emitters
from the x--ray end of the electromagnetic spectrum down to radio waves) is
between 20 to 40 times larger than the luminous matter,
$20\Omega_{lum}\lesssim\Omega_{DM}=\rho_{DM}/\rho_c\lesssim 40\Omega_{lum}$
\item at galactic scales the density of matter in units of the critical
density is such that $\Omega_G=\rho_G/\rho_c\lesssim 0.18$, and \item for
scales up to a few Mpc the density of matter (luminous + dark) is of the
order of 40\% of $\rho_c$ ($\Omega_{DM+Lum}\approx 0.4$).  \end{enumerate}
Further evidence to bear in mind comes from big bang nucleosynthesis (BBN)
which succesfully accounts for the abundances of the primordially
synthesized light nuclei~\cite{bbn}.  BBN places un upper limit for the
present baryon abundance in the Universe at $\Omega_B\lesssim 0.16$.  This
bound and that of point 1. above indicate that most of the baryons in the
Universe could well be dark.  Furthermore the BBN bound together with point
3. suggest that at galactic scales most of the dark matter might be
baryonic.

Baryonic candidates are not in shortage and they are extensively analyzed
in Ref.\cite{olive}.  Experiments aimed at looking at these massive compact
halo objects (MACHO's) are currently under way~\cite{moscoso}.
Concurrently, the search for non--baryonic dark matter, either directly or
indirectly, has already produced some results.  Thus, in Fig.~\ref{fig1},
we show the exclusion plot (at 95\% C.L.) in the cross--section versus
mass--of--the--candidate plane , obtained with two semiconductor detectors
of very low background, the COSME II detector at Canfranc (Spain) and the
TWIN detectors at Homestake (USA).  These detectors have already imposed
stringent bounds on the type of DM particle one should be expecting.  In
the figure we also include the cross section for a heavy Dirac neutrino as
a function of its mass.  Neutrinos are the prototypical non--baryonic
candidates but the list of candidates is not exhausted by them and includes
others, so far, unobserved objects such as:  axions~\cite{turner}, the
lightest supersymmetric partner (neutralino)~\cite{alfaro} and particles
obtained from grand unified theories with unbroken U(1)
symmetries~\cite{hall}. Semiconductor detectors generate a signal by
collecting either the ionization, the photons or the phonons produced by
the detector atoms recoiling after being struck by the particle.  Because
of this process they are best suited to detect candidates that scatter
elastically and coherently from the components of the detector with masses
near the mass of the detector atoms, where the energy deposited by the
candidate will be the largest.  This type of candidates, with masses a few
GeV and larger and with interactions weaker than strong and
electromagnetic, go usually under the generic name of weakly interacting
massive particles (WIMP's).  A typical example is the aforementioned heavy
Dirac neutrino which we use next to illustrate the expected interaction
cross section. A Dirac neutrino with vector couplings to $Z$ bosons and
which scatters elastically from nuclei by $Z$ exchange has a cross section
given by
\begin{equation}
\frac{d\sigma}{dT}=\frac{G^2}{8\pi}\frac{M_R^2}{T_{max}}
\left[Z(1-4\mbox{sin}^4\theta_W)-N\right]^2\underbrace{\mbox{exp}
\left\{-\frac{2M_NTR^2}{3\hbar^2}\right\}}_{\mbox{loss of coherence}}
\label{xsect}
\end{equation}
with $T_{max}$ the maximum recoil energy of the target nucleus of mass
$M_N$, $M_R$ is the reduced mass, $Z$ and $N$ the number of protons and
neutrons in the target nucleus, respectively, and $\theta_W$ the weak
mixing angle.  $G^2=G_F^2(G_f/G_w)^2$ is a generic coupling constant
arbitrarily scaled and which for neutrinos satisfies $G_f=G_w$;
$G_F$ is the Fermi coupling
constant ($G_F^2\approx (290 {\rm GeV})^{-1}=5.24\times 10^{-38}$
cm$^{-2}$).  For comparison we note that, for neutral technibaryons or
strongly coupled $U(1)'$ $(G_f/G_w)^2\approx 10$; for sneutrinos
$(G_f/G_w)^2\approx 2$, and for neutralinos $(G_f/G_w)^2\approx
10^{-1}-10^{-3}$.  The {\em loss of coherence} factor in (\ref{xsect})
takes into account the depletion in the elastic cross section due to
scattering processes where the momentum transfer is larger than the
inverse radius of the target nucleus.

The detection rate, integrated in energy, is given by the product
\begin{equation}
\mbox{Rate}= {\cal N}\times \Phi\times\sigma
\end{equation}
where ${\cal N}$ is the number of target nuclei, $\Phi$ is the dark
matter flux
($\Phi=n<v>$ with $n$ the number density of DM particles and $<v>$ their
mean velocity) and $\sigma$ is the integrated cross section. This
cross section for a particular bin of deposited energy in the detector
$E_{dep}$ has to be convoluted with the velocity distribution of DM
particles in the Earth. Thus, the rate of detected events per energy
interval is expressed in the form
\begin{equation}
R_{E_{dep}}={\cal N}N\times\underbrace{\frac{\rho_{halo}}{m_\delta}}_n
\times\lim_{\Delta T\rightarrow 0}\left\{\Delta T \int_{v_{min}
(m_\delta,T)}^{v_{max}} f(v) v\frac{d\sigma(v,T)}{dT} dv \right\} \label{rate}
\end{equation}
where the three unknown parameters to be determined are:  the halo
density $\rho_{halo}$, the DM particle mass $m_\delta$, and the strength
$G$ of the interaction cross section $d\sigma /dT$ (see
Eq.(\ref{xsect})).  To this rate one has to apply a relative efficiency
factor (REF) which is a function of $E_{dep}$ and depends on the type of
detector used.

Direct searches typically employ the signal--to--noise method in order to
place bounds on potential candidates and their interactions.  The basic
idea is to calculate the ratio of observed to expected rates and to
exclude, for a given candidate mass, all cross sections larger than this
ratio according to some chosen statistical criterion. This is shown in
Fig.~\ref{fig2} where actual data from the COSME II detector are shown as a
function of $E_{dep}$.  Three curves are drawn corresponding to Dirac
neutrinos of masses 10 Gev, 100 GeV and 2 TeV.  The 100 GeV mass curve
greatly exceeds the observed rate, whereas the other two could well be
hidden in the background.  Notice that the lightest candidate (dotted line)
being lighter than the target nucleus can only deposit a limited amount of
energy in head--on collisions (near the threshold of the detector).  The
heaviest candidate, on the other hand, deposits energy on a broader
interval.  Thus, for a given detector material, there is an optimum DM
candidate which has the largest detection rate; it corresponds to that with
a mass in the neighborhood of the detector mass.  As described, the
signal--to--noise method is limited by its definition since it can only set
limits but cannot signal the existence of DM particles. {\em The
sensitivity of any direct search would be greatly enhanced by a well
understood modulation of the expected signal}.

Two types of modulation of the signal have been proposed so far:  First,
the {\em annual modulation} due to the relative velocity of the Earth with
respect to the galactic halo.  In this method one expects to observe
differences in the rate of detection originating on the motion of the Earth
around the Sun which, at some time of the year, adds to the orbital motion
of the Sun about the center of the galaxy ($\vec{V}_{Sun}\approx 250 \pm
25$ km/s plus a peculiar component of $\approx 16.5$ km/s) and six months
later subtracts from the same velocity ($\vec{V}_{Earth}\approx 30$ km/s).
The predicted modulation, of the order $\approx 4-6$\%, affects detection
rates and energy spectra.  Second, {\em diurnal modulation} due to the
elastic scattering of the cold DM particles on the constituent nuclei of
the Earth.  In this method for some values of the masses and coupling
constants of the CDM particles sizable effects in detection rates and
energy spectra are anticipated so long as the detector is placed in a
favorable geographical location.  This last method, originally proposed by
Collar and Avignone~\cite{collar1} in 1992, is the underlying idea behind
the Sierra Grande experiment, that we describe next.

\section{Diurnal modulation and the Sierra Grande experiment}

Figure~\ref{fig3} depicts schematically the Earth crossed out by two axes;
a polar axis $\widehat{Z}$ of rotation about itself, and a translation axis
$\widehat{V}$ denoting the motion of the planet on the galactic plane
following the Sun.  The dark matter halo is assumed at rest in the galactic
coordinate system with a Maxwellian distribution of velocities for its
components, and a dispersion velocity $\vec{v}_{disp}$ of approximately 270
km/s.  A cutoff at the tail of the distribution, given by the escape
velocity from the galaxy $\vec{v}_{esc}\approx 650$ km/s~\cite{drukier} is
also assumed.  For a detector on the Earth one has to add to the velocity
distribution of the DM particles the component due to the motion of the
planet around the center of the galaxy.  This defines naturally a set of
rings (isodetection rings) on the surface of the Earth along which the CDM
flux and velocity distribution will be the same.  As the Earth rotates
about $\widehat{Z}$, a detector located at a given point on it will rotate
through many values of the angle $\theta$ in the figure, and hence the
counting rate and velocity distribution will be modulated according to how
much the elastic scattering of the WIMPs on nuclei in the mantle and core
affect their exit points and velocities.  This will depend on their masses
and coupling constants.

If the scattering of the DM particle with the Earth is considerable, the
greater modulation would be achieved by a detector that manages to get
close to the $\theta\approx 180^0$ isodetection ring where the
probability of at least one scattering is the largest.  Calculations by
Collar and Avignone~\cite{collar2} indicate that such optimal geographic
location corresponds to a detector sitting between 35 and 40$^0$ South
latitude.

An idea about the angles swept by the proposed experiment at Sierra
Grande and other laboratories with similar detectors can be gleaned by
looking at Fig.~\ref{fig4}.  Three sites with detectors in the northern
hemisphere, Canfranc in the border between Spain and France, Homestake
in the USA and Baksan in Siberia, Russia, reach in the optimum case to
$\theta=100^0$.  The Sierra Grande experiment, on the other hand, begins
at $\theta=80^0$ and reaches up to $\theta\approx 180^0$.  Even in the
absence of scattering the geometry depicted in Fig.~\ref{fig3} gives
rise to a modulation of the incoming and outgoing fluxes with respect to
$\theta$ in the way shown in Fig.~\ref{fig5}.  In this situation, a
detector sensitive to events coming in or out of the Earth could well be
used to identify CDM particles.  Associated with the incoming or
outgoing fluxes there is a mean velocity of the DM particles , which
follows a pattern similar to the one shown above (Fig.~\ref{fig6}).

Next step we consider the more realistic situation of particles
penetrating a sphere with a known density profile.  We choose a value of
the incoming flux $\Phi_{in}$ and the incoming velocity $\vec{v}_{in}$
for a very large number of trials, and follow the trajectory of each
particle through the Earth as it scatters and looses energy.  For this,
we model the Earth (see Fig.~\ref{fig7}) according to its elemental
content and geological data and obtain values for the outgoing flux
$\Phi_{out}$ and the outgoing velocity $\vec{v}_{out}$ at each
isodetection ring $\theta$.  Once this is done we recompute the
modified values of $n$ and $f(v)$ of Eq.(\ref{rate}) and calculate the
expected rate for each ring $R(\theta)$.  One such calculation is shown
in Fig.~\ref{fig8} for a candidate with $m_\delta=7$ GeV and a coupling
constant $G_f=31 G_w$.  To get an idea of the expected scattering
probability we notice that for the average density of the Earth\[
<\rho>=1.82\times 10^{23}\mbox{nuclei/cm}^3\] and the average cross
section \[<\sigma>\approx 10^{-32}\mbox{cm}^2\] the probability that a
DM scatters from one of the Earth constituents is \begin{equation}
P=1-\exp\{-2R_\oplus <\rho><\sigma>\}\approx 0.9 \end{equation}

For the realistic case considered above the advantages of placing a
detector in the southern hemisphere become clear by looking at
Fig.~\ref{fig9}. In part $a)$ of the figure we show the expected flux of
DM particles (obtained by Monte Carlo) divided by the expected flux,
for several different masses and coupling constants. The shaded areas on
top of the figure illustrate the $\theta$--bins one could use to compare
detection rates. The experiment at Canfranc sees a very smooth variation
in the flux after crossing all available $\theta$--bands. The one in
Sierra Grande, however, sweeps across those values of $\theta$ where the
modulation is at its largest. The same reasoning applies to the bottom
part of the figure where we show the mean velocity of particles reaching
the detector as a function of the angle $\theta$.

\subsection{The experiment in Sierra Grande}

To search for diurnal modulation effects as described in the previous
section a collaboration between the laboratory TANDAR, Argentina, the
University of South Carolina and Pacific Northwest Laboratory in the
USA, and the University of Zaragoza in Spain, has been set up.  The site
of the experiment is an iron mine (not active since mid 1991) located in
Sierra Grande, province of Rio Negro, some 1,250 km south of Buenos
Aires, Argentina at 41$^0$ South 66$^0$ West.  The mine, that though
inactive has not been abandoned, posseses a central shaft that reaches a
depth of 420 meters with an accompanying downward gallery that spirals
all the way down to the bottom of the mine.  Further 70 km of horizontal
galleries are intended for the exploitation of the mineral.  Formerly a
state company, the provincial administration which now runs the mine has
granted for the experiment an easy accesible area at approximately 360 m
deep ($\approx 1,000$ mwe) where the detector and all accesory equipment
will be installed.

The planned detector is a 1 kg hyperpure crystal of natural Germanium
($\gtrsim$92\% spin--zero nuclei) built by Princeton Gamma--Tech.  The
cryostat and related low background technology has been developed by the
people at Pacific Northwest Laboratory, Richland, Seattle, and the whole
setup is similar to others used by the same group for double beta decay
and DM experiments in different laboratories.  The data taking will be
the standard of a nuclear physics experiment and will be performed on
what is called an event by event basis, namely for each event one
records its energy and time tags its arrival in order to associate the
time with the corresponding angle $\theta$.  When enough statistics is
collected, for each energy bin $i$, the data are summed which correspond
to an interval in angle where the counting rate was low $\Delta\theta_2$
and subtracted from the collected data at an angle region where the
counting rate was high $\Delta\theta_1$.  The difference in the number
of events between these two $\theta$--regions $R_i= N(\Delta\theta_2)
-N(\Delta\theta_1)$ is called the {\em residual} for energy bin $i$; if
a statiscally significant deviation from zero shows up in $R_i$ it will
be indicating the presence of a DM particle in a limited region of the
$(m_\delta,\:  G)$ plane, the precise value depending on the statistics
collected.  One nice feature of this sort of analysis is that since
there is a subtraction involved, the statistical error bars decrease as
more data are considered whereas in the signal--to--noise method the
errors bars are, at best, as good as those of the unknown background.

In Fig.~\ref{fig10} we show a Monte Carlo residuals spectrum for a 1 TeV
heavy Dirac neutrino assuming a background similar to those already
achieved by existing detectors.  The dashed curve is the theoretical
result.  One point to mention that in all calculations the density of
the halo has been kept fixed at $\rho_{halo}=0.4 $ GeV
cm$^{-3}$ which is the most often quoted value.  Uncertainties in this
number, which have been risen~\cite{kuijken} (specially towards smaller
densities), will move upwards the excluded region in Fig.~\ref{fig1}
and make this sort of analysis even more relevant.

\section{Conclusions}

For some ranges of masses and coupling constants, CDM particles can
suffer significant scattering with the Earth that could result in
experimentally observable modulation in existing detectors. Subtracting
energy spectra for different times of the day increases the sensitivity
for placing bounds on certain CDM candidates. A ``real effect" would
result in a positive signal over the nature ($m_\delta,\: G$) of CDM
particles. The planned experiment in Sierra Grande--Argentina is the
first attempt to exploit this diurnal modulation effect.

\begin{figure}
\caption{ Exclusion plot (95\% C.L.) for the mass and elastic cross
section on Ge for dark matter particles. The solid line corresponds to
the weak interaction cross sction (heavy Dirac neutrino).\label{fig1}}
\end{figure}
\begin{figure}
\caption{The normalized raw spectra from the COSME II detector presently
operated by the UZ/PNL/USC collaboration. The expected detection rates
for different heavy Dirac neutrinos in the galactic halo are also shown
(ref.[11]).\label{fig2}}
\end{figure}
\begin{figure}
\caption{The geometry of the diurnal modulation method; shown are the
velocity of the Earth through the galactic halo, $\widehat{V}$, the
Earth's axis of rotation $\widehat{Z}$, and the isodetection rings on
which the CDM distribution and flux is uniform
(ref.[12]).\label{fig3}}
\end{figure}
\begin{figure}
\caption{Bands in the angle $\theta$ of Fig.~3 swept by three different
detectors in the northern hemisphere, and by the proposed experiment in
Sierra Grande (from [12]).\label{fig4}}
\end{figure}
\begin{figure}
\caption{Incoming and outgoing fluxes as function of the isodetection
rings location (from  [12]). No scattering with the Earth.\label{fig5}}
\end{figure}
\begin{figure}
\caption{Similar to Fig.~5 but plotting the mean velocity as function of
$\theta$.\label{fig6}}
\end{figure}
\begin{figure}
\caption{Cross section of the density profile of the Earth indicating
the relative abundances. \label{fig7}}
\end{figure}
\begin{figure}
\caption{Comparison of the normalized fluxes into and out of the Earth
without scattering with the flux out and total flux with scattering in
the Earth.  The chose CDM candidate has a mass $m_\delta=7$ GeV and
$G_f^2= 1000\:G_w^2$. \label{fig8}}
\end{figure}
\begin{figure}
\caption{a) Sample calculations of the flux of acttered CDM particles
normalized to the flux without scattering for 4 values of mass and
coupling constant. b) same as a) but for mean velocities. In both
figures the regions in $\theta$ crossed by the COSME II detector
and the Sierra Grande experiment are shown.\label{fig9}}
\end{figure}
\begin{figure}
\caption{Simulated experimental residual for CDM consisting of heavy
Dirac neutrinos. The detector modelled here is COSME II which cannot
reject these candidates using conventional exclusion
techniques.\label{fig10}}
\end{figure}

\end{document}